\documentclass[nofootinbib, twocolumn]{revtex4}
\usepackage{amsfonts}
\usepackage{graphicx}
\begin{document}
\title{Variational problem for Hamiltonian system on so(k, m) Lie-Poisson
manifold and dynamics of semiclassical spin}

\author{A. A. Deriglazov} \email{alexei.deriglazov@ufjf.edu.br}
\altaffiliation{On leave of absence from Dep. Math. Phys., Tomsk Polytechnical University, Tomsk, Russia.}

\affiliation{Depto. de Matem\'atica, ICE, Universidade Federal de Juiz de Fora, MG, Brasil}

\begin{abstract}
We describe the procedure for obtaining Hamiltonian equations on a manifold with $so(k, m)$ Lie-Poisson bracket from a
variational problem. This implies identification of the manifold with base of a properly constructed fiber bundle
embedded as a surface into the phase space with canonical Poisson bracket. Our geometric construction underlies the
formalism used for construction of spinning particles in \cite{AAD2, AAD3, AAD7, AAD4}, and gives precise mathematical
formulation of the oldest idea about spin as the "inner angular momentum".
\end{abstract}

\maketitle 

\section{Introduction}
Typical spinning-particle model consist of a point on a world-line
and some set of variables describing the spin degrees of freedom,
which form the inner space attached to that point \cite{HR}. In
fact, different spinning particles discussed in the literature
differ by the choice of inner space of spin. The choice is
dictated by algebra of commutators of spin operators in quantum
theory. This is the algebra of angular momentum. For example, for
the case of non-relativistic spin (Pauli equation), the operators
$\hat S^1, \hat S^2, \hat S^3$ are proportional to the Pauli
matrices and obey $so(3)$ algebra $[\hat S^i, \hat
S^j]=i\hbar\epsilon^{ijk}\hat S^k$. If we intend to arrive at the
algebra starting from a variational problem of classical
mechanics, the most natural way is to consider the spin variables
as the composed quantities, $S^i=\epsilon^{ijk}\omega^j\pi^k$,
where $\omega, \pi$ are coordinates of phase space equipped with
canonical Poisson bracket. Unfortunately, this is not the whole
story. First, we need some mechanism which explains why $\vec S$,
not $\omega$ and $\pi$ must be taken for the description of spin
degrees of freedom. Second, the basic space is six-dimensional,
while the spin manifold is two-dimensional (remind that the square
of spin operator has fixed value, $\hat{\bf
S}{}^2=\frac{3\hbar^2}{4}$). To improve this, we need to impose
constraints on the basic variables. This implies the use of Dirac
machinery for constrained theories.

Following these lines, various non-Grassmann spinning-particle
Lagrangians have been constructed and analyzed in the recent works
\cite{AAD2, AAD3, AAD7, AAD4, AAD5}. In \cite{AAD2} it has been
demonstrated that $so(3)$ algebra leads to a reasonable model of
non-relativistic spin. $so(1, 3)$ algebra can be used to construct
variational problem for unified description of both the Frenkel
\cite{Fre} and BMT \cite{BMT} theories of relativistic spin
\cite{AAD3}. $so(2, 3)$ algebra implies two different models
associated with the Dirac equation \cite{AAD4, AAD7, AAD5}.

In the present work we describe the unique geometric construction
of spin surface which lies behind the models. We discard the
world-line variables and concentrate on the structure of spin
sector of the models (the case of frozen spin). This reduces the
problem to that of formulation of a variational problem for
Hamiltonian system on a manifold with Lie-Poisson bracket.

We analyze and solve the problem for $so(k, m)$ Lie-Poisson manifold, the case which has immediate applications, as we
have mentioned above. Spin surface will be identified with base of spin fiber bundle determined by the same system of
algebraic equations in any dimension. On the geometric language, this is the structure of fiber bundle that forces us
to describe the spin degrees of freedom by the angular-momentum coordinates.

Hamiltonian systems on Poisson manifolds naturally arise during analysis of many classical problems \cite{Arn, Per,
Olv} and in modern extensions and applications of Hamiltonian formalism \cite{AN1, GEN1, GEN3, GEN4, GEN5, AAD8, GEN2}.
Numerous examples of dynamics on nontrivial Poisson manifolds can be obtained applying the Dirac procedure for analysis
of constrained systems to singular Lagrangian theories \cite{AN1, GEN0, OTH1, GEN3, AAD10}. So, the inverse task we
address in this work represents certain interest on its own right. Having at hands the variational problem, we would be
able to carry out more systematic and unequivocal analysis of the models under intensive study in various branches of
current interest \cite{NUC1, NUC2, DSR1}, including non commutative geometry \cite{NCG1, AAD10, NCG2}.

Having in mind physical applications, we use the local
coordinates. Conversion to coordinate free setting will be
reported elsewhere.

The work is organized as follows. To formulate variational problem for $so(k, m)$ Lie-Poisson bracket, we embed the
spin surface into a properly constructed phase space with canonical Poisson bracket. The embedding procedure described
in section 2, and leads to identification of spin surface with a base of spin fiber bundle. Its structure group
described in details in section 3. Since the embedding can be treated as imposition of constraints on phase space, in
sections 4, 5 we look for the action functional which generates the desired constraints. Both Hamiltonian and
Lagrangian actions are found in closed form. Some technical details collected in the Appendices A and B.

\section{Spin surface and associated spin fiber bundle}

The formulation of a variational problem in closed form is known
for Hamiltonian system defined on phase space with canonical
Poisson bracket, $\{\omega_i, \pi_j\}=\delta_{ij}$. Let $H(\omega,
\pi)$ stands for Hamiltonian of the system. Then Hamiltonian
equations can be obtained by variation of the action
\begin{eqnarray}\label{so.0.1}
\int d\tau ~ \pi\dot\omega-H(\omega, \pi).
\end{eqnarray}

Consider the same problem on symplectic manifold, that is
$2n$\,-dimensional manifold with local coordinates $z^a$, endowed
with a closed nondegenerate differential 2-form
$\omega^{(2)}_{ab}(z)dz^a\land dz^b$. This determines the bracket
$\{z^a, z^b\}=\omega_{(2)}^{ab}$, where $\omega_{(2)}$ is inverse
matrix of $\omega^{(2)}$. This case reduces to the previous one.
According to Darboux's theorem, we can pass from $z^a$ to the
canonical coordinates $\omega^i, \pi^j$, $z^a=z^a(\omega, \pi)$.
Then Hamiltonian equations for $\omega$ and $\pi$ follow from
(\ref{so.0.1}) with $H(\omega, \pi)\equiv H(z^a(\omega, \pi))$.
Evolution of the initial variables reads
$z^a(\tau)=z^a(\omega(\tau), \pi(\tau))$.

Poisson manifold represents more general case, when the structure
function $\omega_{(2)}$ does not supposed to be invertible (this
includes the case of odd-dimensional manifold). In particular,
Lie-Poisson bracket is defined by
$\omega_{(2)}^{ij}=c^{ij}{}_kz^k$, where $c^{ij}{}_k$ are
structure constants of a Lie algebra. This is the case we discuss
in the present work. We consider $\frac12n(n-1)$-dimensional space
with the metric $\eta=(-, \ldots, -, +, \ldots , +)$, equipped
with the coordinates $J^{\mu\nu}=-J^{\nu\mu}$, $\mu, \nu=0, 1, 2,
\ldots , n=k+m$ and with the Lie-Poisson bracket
\begin{eqnarray}\label{so.1}
\mathbb{R}^{\frac{n(n-1)}{2}}=\Biggl\{ J^{\mu\nu}, ~  \{J^{\mu\nu}, J^{\alpha\beta}\}_{LPB}= \cr
2(\eta^{\mu\alpha}J^{\nu\beta}-\eta^{\mu\beta}J^{\nu\alpha}-
\eta^{\nu\alpha}J^{\mu\beta}+\eta^{\nu\beta}J^{\mu\alpha})\Biggr\}.
\end{eqnarray}
This is the Lie-Poisson manifold associated with $so(k, m)$ algebra \cite{Per, Olv}.

We discuss the Hamiltonian flow
\begin{eqnarray}\label{so.0}
\dot J=\{J, H\}_{LPB},
\end{eqnarray}
generated by given Hamiltonian $H(J)$. Our aim is to formulate
variational problem for the Hamiltonian flow on the submanifold
$\mathbb S$ which will be specified below. We call $\mathbb S$
spin surface, as the canonical quantization of the submanifold
gives quantum mechanics of spin one-half particle.

Our first task is to generate Lie-Poisson bracket starting from
the canonical Poisson bracket. To achieve this, we use vector
representation of $so(k, m)$ (more generally, any linear
representation can be used to this aim, see Appendix A for
details).

Consider $2n$-dimensional phase space equipped with the Poisson bracket
\begin{eqnarray}\label{so.2}
\mathbb{R}^{2n}=\{~ \omega^\mu, \pi^\nu, ~ \{\omega^\mu,
\pi^\nu\}_{PB}=\eta^{\mu\nu} ~ \}.
\end{eqnarray}
Define the map from the phase to angular-momentum space (\ref{so.1})
\begin{eqnarray}\label{so.3}
f: ~ \mathbb{R}^{2n} ~ \rightarrow ~ \mathbb{R}^{\frac{n(n-1)}{2}}, \qquad \qquad \quad \qquad ~ \cr f: (\omega^\mu,
\pi^\nu) ~ \rightarrow ~ J^{\mu\nu}=2(\omega^\mu\pi^\nu-\omega^\nu\pi^\mu). ~
\end{eqnarray}
We have, for $n>2$,
\begin{eqnarray}\label{so.4}
\mbox{rank}\frac{\partial(J^{\mu\nu})}{\partial(\omega^k, \pi^l)}=2n-3,
\end{eqnarray}
so an image of the map is $(2n-3)$-dimensional surface $\mathbb{M}$
\begin{eqnarray}\label{so.5}
f(\mathbb{R}^{2n})=\mathbb{M}^{2n-3}\in\mathbb{R}^{\frac{n(n-1)}{2}}.
\end{eqnarray}
Poisson bracket of the functions $J^{\mu\nu}(\omega, \pi)$ coincides with the Lie-Poisson bracket (\ref{so.1}). More
generally, for any functions $A(J)$, $B(J)$,
\begin{eqnarray}\label{so.5.1}
\{A(J), B(J)\}_{PB}=\left.\{A(J), B(J)\}_{LPB}\right|_{J\rightarrow J(\omega, \pi)}.
\end{eqnarray}

Further, to improve wrong balance of degrees of freedom (see Eq. (\ref{so.4})), we look for the surface $\mathbb{T}=\{
~ \omega, \pi ~ | ~ T_a(\omega^\mu, \pi^\nu)=0 ~ \}\in \mathbb{R}^{2n}$ which is invariant under action of $SO(n)$,
that is
\begin{eqnarray}\label{so.6}
\{T_a, J^{\mu\nu}\}_{PB}=0.
\end{eqnarray}
There is essentially unique invariant surface of $2n-3$ dimensions
\begin{eqnarray}\label{so.7}
\mathbb{T}^{2n-3}=\{ T_3=\pi^2+a_3=0, ~  T_4=\omega^2+a_4=0, \cr
 T_5=\omega\pi+a_5=0 \}, \qquad \qquad
\end{eqnarray}
where $a_3, a_4, a_5\in\mathbb R$, and it has been denoted
$\pi^2=\pi^\mu\pi_\mu$, and so on.
\par

\noindent {\it Comments.} A.  Any trajectory of $H(J)$ which starts on $\mathbb{M}^{2n-3}$ lies entirely on
$\mathbb{M}^{2n-3}$ (the proof is similar to those of Proposition 3 below).\par

\noindent B. $so(3)$ is the exceptional case, when $\mathbb
M=\mathbb{R}^{\frac{n(n-1)}{2}}$, and the vector representation of
$so(3)$ coincides with the adjoint one. Besides, the surface
$\mathbb{T}^{2n-3}$ can be identified with the group manifold
$SO(3)$, see \cite{AAD6} for details. \par

\noindent C. The invariance condition (\ref{so.6}) guarantees the validity of important Propositions 2, 3, see below.
\par \noindent D. Casimir operators of $SO(n)$ group
are scalar functions of generators, $C(J^{\mu\nu})$. On the surface (\ref{so.7}) they have fixed values determined by
the constants $a_3$ and $a_4$: $C(J^{\mu\nu})=C(\omega^2, \pi^2, \omega\pi)=C(a_3, a_4, a_5)$. In particular, the first
Casimir operator is $J^2=8[\omega^2\pi^2-(\omega, \pi)^2]=8[a_3a_4-a_5^2]$.

Denote $\mathbb{S}$ image of $\mathbb{T}^{2n-3}$ under the map $f$ (this is called the spin surface, see Figure 1)
\begin{eqnarray}\label{so.8}
\mathbb{S}=f(\mathbb{T}^{2n-3})\in\mathbb{R}^{\frac{n(n-1)}{2}}.
\end{eqnarray}
Denote $\mathbb{F}_J\in\mathbb{T}^{2n-3}$ preimage of a point
$J\in\mathbb{S}$, $\mathbb{F}_J=f^{-1}(J)$.

\begin{figure*}[t] \centering
\includegraphics[width=420pt, height=250pt]{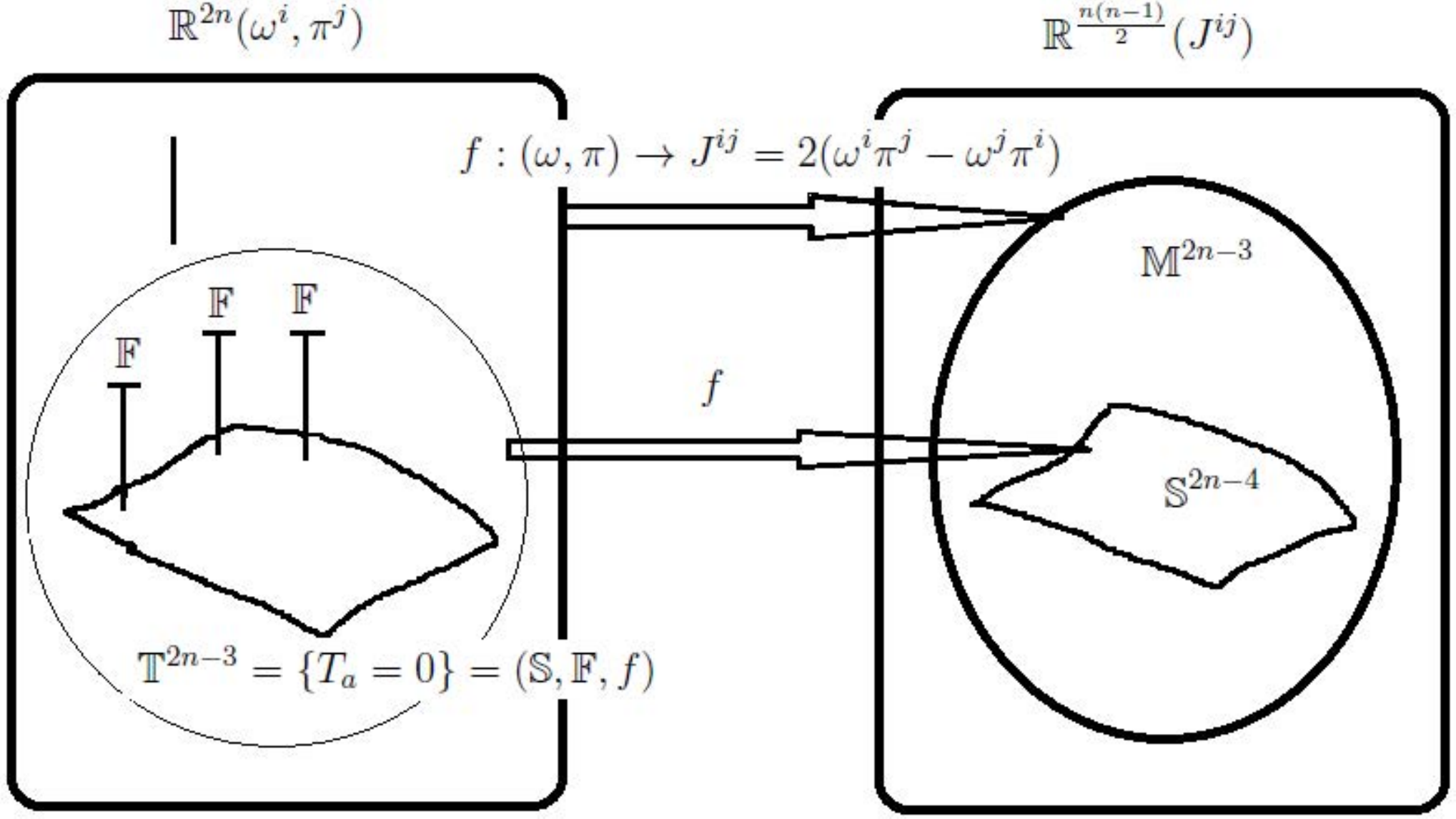}
\caption{Identification of spin surface $\mathbb{S}^{2n-4}$ with base of the spin fiber bundle
$\mathbb{T}^{2n-3}$.}\label{Fig}
\end{figure*}

Then the manifold $\mathbb{T}^{2n-3}$ acquires natural structure of fiber bundle
\begin{eqnarray}\label{so.9}
\mathbb{T}^{2n-3}=(\mathbb{S}, \mathbb{F}, f),
\end{eqnarray}
with the base $\mathbb{S}$, the projection map $f$, the standard fiber $\mathbb{F}$. Structure group of the fiber will
be described in section 3.

Local coordinates on $\mathbb{M}^{2n-3}$ and equations of this surface can be obtained solving Eq. (\ref{so.3}).
Namely, the subset $J'$ of $2n-3$ independent functions among $J(\omega, \pi)$, represents the local coordinates.

Let us discuss this in some details. In accordance with the rank condition (\ref{so.4}), we can separate  $J^{\mu\nu}$
on two groups, $J=(J', J'')$, in such a way that the number of $J'$ is equal to $2n-3$, and
\begin{eqnarray}\label{so.10}
\mbox{rank}\frac{\partial J'}{\partial(\omega, \pi)}=2n-3.
\end{eqnarray}
Then equations for $J'$ of the system (\ref{so.3}) can be resolved
with respect to some of $2n-3$ variables among $\omega$ and $\pi$.
Substitute these expressions into the remaining equations from
(\ref{so.3}). By construction, the result does not depend on
$\omega$ and $\pi$, so we obtain expressions for $J''$ through
$J'$
\begin{eqnarray}\label{so.11}
J''=g(J')\equiv J''(J').
\end{eqnarray}
In the result, the image $f(\mathbb{R}^{2n})$ is the surface with
equations (\ref{so.11}) and with local coordinates $J'$
\begin{eqnarray}\label{so.12}
\mathbb{M}^{2n-3}=\Biggl\{J=(J', J'') ~ \Big| ~ J''=J''(J')\Biggr\}.
\end{eqnarray}
So, points of $\mathbb{R}^{2n}$ are mapped into
\begin{eqnarray}\label{so.13}
f(\omega, \pi)=(J', J''(J'))\in\mathbb{M}^{2n-3}.
\end{eqnarray}

For example, for $SO(2, 3)$ case, the local coordinates are
$J^{5\mu}$, $J^{0i}$ while as the equations of the surface we can
take \cite{AAD5}
\begin{eqnarray}\label{so.14}
\epsilon^{\mu\nu\alpha\beta}J^5{}_\nu J_{\alpha\beta}=0,
\Leftrightarrow J^{ij}=(J^{50})^{-1}(J^{5i}J^{0j}-J^{5j}J^{0i}).
\end{eqnarray}
Among the four relativistic-covariant equations on the l.h.s.,
there are only three independent. \vspace{3ex} \par

In a vicinity of each point of $\mathbb{T}^{2n-3}$ we can choose
local coordinates adjusted with the structure of fibration. That
is we look for the coordinates $(\tilde J, \tilde\omega)$, where
the group $\tilde J$ parameterize the base $\mathbb{S}$ while
$\tilde\omega$ parameterize the fiber $\mathbb{F}$. We observe
that
\begin{eqnarray}\label{so.15}
\mbox{rank}\frac{\partial(J',\omega^n, T_4,
T_5)}{\partial(\omega^k, \pi^l)}=2n,
\end{eqnarray}
so we can make the following change of coordinates on
$\mathbb{R}^{2n}$,
\begin{eqnarray}\label{so.16}
(\omega^i, \pi^j) ~ \leftrightarrow ~ (J', \omega^n, T_4, T_5).
\end{eqnarray}
In the new coordinates the function $T_3$ does not depend on $\omega^n$. Indeed, $J^{\mu\nu}J_{\mu\nu}$ can be
identically rewritten through $T_a$ as follows
\begin{eqnarray}\label{so.17}
J^2=8\left[(T_4-a_4)T_3-(T_5-a_5)^2-a_3T_4+a_3a_4\right],
\end{eqnarray}
then
\begin{eqnarray}\label{so.18}
T_3=\frac{J^2+8(T_5-a_5)^2+8a_3T_4-8a_3a_4}{8(T_4-a_4)}.
\end{eqnarray}
On the other hand, substitute the new coordinates into the expression $(J^{\mu\nu}(\omega, \pi))^2$. By construction,
this gives $J^2=(J')^2+(J''(J'))^2$. Using this in Eq. (\ref{so.18}), we obtain
\begin{eqnarray}\label{so.19}
\left. T_3(\omega, \pi)\right|_{(J', \omega^n, T_4, T_5)}=T_3(J',
T_4, T_5).
\end{eqnarray}
Hence in the new coordinates the surface looks as
\begin{eqnarray}\label{so.20}
\mathbb{T}^{2n-3}=\Biggl\{J', \omega^n, T_4, T_5 ~ \Big| ~
T_4=T_5=0, ~  T_3(J')=0 \Biggr\}.
\end{eqnarray}
{\it Proposition 1.} $\dim\mathbb{S}=2n-4$, ~ (then $\dim\mathbb{F}=1$).

Indeed, take restriction of the map (\ref{so.13}) on
$\mathbb{T}^{2n-3}$. Since any point on the surface obeys the
condition $T_3(J')=0$, we have
\begin{eqnarray}\label{so.21}
\left. f\right|_{\mathbb T}: (\omega, \pi) ~ \rightarrow ~ (J',
J''(J')), \quad \mbox {where} \quad T_3(J')=0.
\end{eqnarray}
Hence $T_3(J')=0$ is equation of the base in space
$\mathbb{M}^{2n-3}$, and $\dim\mathbb S=2n-4$.

Let as take some $2n-4$ variables $\tilde J$ among $J'$ which form
a coordinate system of the base $\mathbb{S}^{2n-4}$. Then
(\ref{so.16}) and (\ref{so.20}) imply that $(\tilde J, \omega^n)$
can be taken as local coordinates of the fibration
$\mathbb{T}^{2n-3}$. In the dynamical model constructed in section
3, the coordinates $\tilde J$ represent observable quantities
while the coordinate $\omega^n$ is pure gauge degree of freedom.

\section{Structure group and spin-plane local symmetry}
Structure group of the standard fiber turn into the local symmetry
of dynamical theory. So it is important to find manifest form of
the transformations.

For the case of Euclidean space, $so(n)$, we have $a_3<0$, $a_4<0$ and $a_3a_4-a_5^2\ne 0$. The last condition
guarantees $J^2\ne 0$, the case which we are interested in. Let us identify the point $(\omega,
\pi)\in\mathbb{T}^{2n-3}$ with pair of vectors of Euclidean space $\mathbb{R}^n$
\begin{eqnarray}\label{so.9.1}
\vec\pi^2=-a_3, \quad \vec\omega^2=-a_4, \quad (\vec\omega, \vec\pi)=-a_5.
\end{eqnarray}
Given point $(\omega, \pi)$ with
$J^{\mu\nu}=2\omega^{[\mu}\pi^{\nu]}$, the fiber
$\mathbb{F}_J=f^{-1}(J)$ is composed by all the pairs obtained
from $(\vec\omega, \vec\pi)$ by rotations in the plane of these
vectors, see Appendix B for the proof. Denote $\frac{(\vec\omega,
\vec\pi)}{|\vec\omega||\vec\pi|}=\cos\sigma$, then the rotation on
angle $\beta$ reads
\begin{eqnarray}\label{so.21.1}
\vec\omega'=-\vec\omega\frac{\sin(\beta-\sigma)}{\sin\sigma}+\vec\pi\frac{|\vec\omega|\sin\beta}{|\vec\pi|\sin\sigma},
\cr \vec\pi'=-\vec\omega\frac{|\vec\pi|\sin\beta}{|\vec\omega|\sin\sigma}+\vec\pi\frac{\sin(\beta+\sigma)}{\sin\sigma}.
\end{eqnarray}
By construction, $(\omega, \pi)\in\mathbb{T}^{2n-3}$ implies $(\omega', \pi')\in\mathbb{T}^{2n-3}$. Infinitesimal form
of the symmetry is
\begin{eqnarray}\label{so.21.2}
\delta\vec\omega=\beta\vec\pi\frac{|\vec\omega|}{|\vec\pi|\sin\sigma}-\beta\vec\omega\cot\sigma, \cr
\delta\vec\pi=-\beta\vec\omega\frac{|\vec\pi|}{|\vec\omega|\sin\sigma}+\beta\vec\pi\cot\sigma.
\end{eqnarray}

For the case $so(k, m)$, manifest form of transformations depend on the values of constants $a_i$.

1. If $\omega^2$ and $\pi^2$ have different signs, then $-\infty<\frac{(\omega\pi)^2}{\omega^2\pi^2}<0$, so there is
$\sigma$ such that $\frac{(\omega\pi)^2}{\omega^2\pi^2}=-\sinh^2\sigma$. The transformations which leave invariant
$\omega^2$, $\pi^2$, $(\omega\pi)$ and $J^{\mu\nu}$ are
\begin{eqnarray}\label{sp.1}
\omega'^\mu=\frac{1}{\cosh\sigma}\left(\omega^\mu\cosh(\beta-\sigma)+
\pi^\mu\frac{\omega^2}{(\omega\pi)}\sinh\beta\sinh\sigma\right), \cr
\pi'^\mu=\frac{1}{\cosh\sigma}\left(-\omega^\mu\frac{\pi^2}{(\omega\pi)}\sinh\beta\sinh\sigma
+\pi^\mu\cosh(\beta+\sigma)\right).
\end{eqnarray}
Infinitesimal form of the transformation is
\begin{eqnarray}\label{sp.2}
\delta\omega^\mu=-\beta\omega^\mu\tanh\sigma+\beta\pi^\mu\frac{\omega^2\tanh\sigma}{(\omega\pi)}, \cr
\delta\pi^\mu=-\beta\omega^\mu\frac{\pi^2\tanh\sigma}{(\omega\pi)}+\beta\pi^\mu\tanh\sigma.
\end{eqnarray}

2. If $\omega^2$ and $\pi^2$ have the same sign, then
$\frac{(\omega\pi)^2}{\omega^2\pi^2}>1$, so there is $\sigma$ such
that $\frac{(\omega\pi)^2}{\omega^2\pi^2}=\cosh^2\sigma$. The
structure group is
\begin{eqnarray}\label{sp.3}
\omega'^\mu=\frac{1}{\sinh\sigma}\left(-\omega^\mu\sinh(\beta-\sigma)+
\pi^\mu\frac{\omega^2}{(\omega\pi)}\sinh\beta\cosh\sigma\right), \cr
\pi'^\mu=\frac{1}{\sinh\sigma}\left(-\omega^\mu\frac{\pi^2}{(\omega\pi)}\sinh\beta\cosh\sigma
+\pi^\mu\sinh(\beta+\sigma)\right).
\end{eqnarray}
Infinitesimal form of the transformation is
\begin{eqnarray}\label{sp.4}
\delta\omega^\mu=-\beta\omega^\mu\coth\sigma+\beta\pi^\mu\frac{\omega^2\coth\sigma}{(\omega\pi)}, \cr
\delta\pi^\mu=-\beta\omega^\mu\frac{\pi^2\coth\sigma}{(\omega\pi)}+\beta\pi^\mu\coth\sigma.
\end{eqnarray}

3. If $\omega^2$ and $\pi^2$ lie on light-cone, $\omega^2=0$, $\pi^2=0$, the structure group is ($\beta\ne 0$)
\begin{eqnarray}\label{sp.5}
\omega'^\mu=\beta\omega^\mu, \qquad \pi'^\mu=\frac{1}{\beta}\pi^\mu.
\end{eqnarray}

4. If $\omega^2=0$ but $\pi^2\ne 0$, $(\omega\pi)\ne 0$ the structure group is ($\beta\ne 0$)
\begin{eqnarray}\label{sp.6}
\omega'^\mu=\frac{1}{\beta}\omega^\mu, \qquad
\pi'^\mu=\frac{(1-\beta^2)\pi^2}{2\beta(\omega\pi)}\omega^\mu+\beta\pi^\mu.
\end{eqnarray}

5. If $\omega^2=0$ and $(\omega\pi)=0$, but $\pi^2\ne 0$,  the structure group is
\begin{eqnarray}\label{sp.7}
\omega'^\mu=\omega^\mu, \qquad \pi'^\mu=\beta\omega^\mu+\pi^\mu.
\end{eqnarray}

By construction, the transformations leave inert points of base, $\delta J^{\mu, \nu}=0$. In the dynamical realization
of section 4, the structure group acts independently at each instance of time and turn into the local (gauge) symmetry
which we call spin-plane symmetry. This determines physical sector of the theory, and hence play the fundamental role
in our construction. Indeed, according to Eq. (\ref{so.3}), we consider the spin $J^{\mu, \nu}$ as angular-momentum of
an "inner-space particle" $\omega \mu$. The crucial difference with the usual (spacial) angular momentum is the
presence of spin-plane symmetry, which acts on the basic variables $\omega$, $\pi$, while leaves invariant the spin
variables $J$. According to the general theory \cite{dir, GT1, AAD1}, the gauge non-invariant coordinates $\omega$ of
the inner-space are not physical (observable) quantities. The only observable quantities are the gauge-invariant
variables $J$. So our geometric construction realizes, in a systematic form, the oldest idea about spin as the "inner
angular momentum".

\section{Variational problem for Hamiltonian system with $so(k, m)$ Lie-Poisson bracket}
Let $H(J)$ is some Hamiltonian on the Lie-Poisson manifold
(\ref{so.1}). The map $f$ can be used to induce the Hamiltonian
$H(\omega, \pi)$ on the phase space $\mathbb{R}^{2n}$
\begin{eqnarray}\label{so.22}
H(\omega, \pi)\equiv H(J(\omega,\pi)).
\end{eqnarray}
Let us confirm that Hamiltonian flows of $H(\omega, \pi)$ and
$H(J)$ are adjusted with the surfaces $\mathbb{T}^{2n-3}$ and
$\mathbb{S}$. \par

\noindent {\it Proposition 2.} Any trajectory of $H(\omega, \pi)$
which starts on $\mathbb{T}^{2n-3}$ lies entirely on
$\mathbb{T}^{2n-3}$. \par

Indeed, let $T_a(\omega(\tau_0), \pi(\tau_0))=0$ for some
trajectory of $H(\omega, \pi)$. We have
\begin{eqnarray}\label{so.23}
\dot T_a(\omega(\tau), \pi(\tau))=\{T_a, H(J(\omega, \pi))\}_{PB}= \cr \{T_a, J\}_{PB}\frac{\partial H}{\partial J}=0,
\end{eqnarray}
due to the invariance condition $\{T_a, J\}=0$. Hence
$T_a(\tau)=T_a(\tau_0)=0$ for any $\tau$, that is $(\omega(\tau),
\pi(\tau))$ belong to $\mathbb{T}^{2n-3}$ at each $\tau$.
\par

\noindent {\it Proposition 3.} Any trajectory of $H(J)$ which starts on $\mathbb{S}$ lies entirely on $\mathbb{S}$.

Indeed, the problem
\begin{eqnarray}\label{so.24}
\dot J=\{J, H(J)\}_{LPB}=0, \qquad J(\tau_0)=J_0\in\mathbb{S},
\end{eqnarray}
has unique solution, we denote it $J(\tau)$. Take any point $(\omega_0,\pi_0)$ which belong to preimage of $J_0$,
$f(\omega_0,\pi_0)=J_0$. Construct the (unique) solution to the problem $\dot\omega=\{\omega, H(\omega, \pi)\}_{PB}=0$,
$\dot\pi=\{\pi, H(\omega, \pi)\}_{PB}=0$, $(\omega(\tau_0), \pi(\tau_0))=(\omega_0, \pi_0)$. According the Proposition
2, it lies in $\mathbb{T}^{2n-3}$, then $f(\omega(\tau), \pi(\tau))$ lies in $\mathbb{S}$. Besides, this obeys the
problem (\ref{so.24}). Since solution to the problem is unique, we conclude $J(\tau)=f(\omega(\tau),
\pi(\tau))\in\mathbb{S}$ for each $\tau$.

We are ready to formulate variational problem for $so(k, m)$ Lie-Poisson system (\ref{so.0}). Consider the action
functional on the extended phase space $(\omega^\mu, \pi^\nu, e_a, \pi_a, \lambda_{ea})$, $a=3, 4, 5$,
\begin{eqnarray}\label{so.25}
S_H=\int d\tau ~  \pi\dot\omega-\left[H(J(\omega,
\pi))+\frac{e_a}{2}T_a+\pi_{ea}(\lambda_{ea}-\dot e_a)\right].
\end{eqnarray}
Variation of the functional leads to the equations
\begin{eqnarray}\label{so.26}
\pi_{ea}=0, \quad \dot e_a=\lambda_{ea},
\end{eqnarray}
\begin{eqnarray}\label{so.27}
\pi^2+a_3=0, \quad \omega^2+a_4=0, \quad (\omega\pi)+a_5=0,
\end{eqnarray}
\begin{eqnarray}\label{so.28}
\dot\omega^\mu=\frac{\partial H}{\partial\pi^\mu}+e_3\pi^\mu+\frac12 e_5\omega^\mu, \cr \dot\pi^\mu=-\frac{\partial
H}{\partial\omega^\mu}-e_4\omega^\mu-\frac12 e_5\pi^\mu.
\end{eqnarray}
Eq. (\ref{so.26}) has been used in obtaining (\ref{so.27}).

Equations (\ref{so.27}) and (\ref{so.7}) imply that all the trajectories $(\omega(\tau), \pi(\tau))$ of the problem
(\ref{so.25}) live on the fiber bundle $\mathbb{T}^{2n-3}$.

We point out that Eq. (\ref{so.22}) implies useful identities
\begin{eqnarray}\label{so.28.1}
\pi^\mu\frac{\partial H}{\partial\omega^\mu}=\omega^\mu\frac{\partial H}{\partial\pi^\mu}=0, \qquad \quad \cr
\omega^\mu\frac{\partial H}{\partial\omega^\mu}=\pi^\mu\frac{\partial H}{\partial\pi^\mu}=\frac{\partial H}{\partial
J^{\alpha\beta}}J^{\alpha\beta}.
\end{eqnarray}
The system (\ref{so.26})-(\ref{so.28}) contains algebraic equations (\ref{so.27}). So we use the Dirac prescription to
deduce all the algebraic consequences of the system. Compute derivative of the first equation from (\ref{so.27}). This
gives $2e_4a_5+e_5a_3=0$. In turn, derivative of this equation determines one of Lagrangian multipliers,
$2\lambda_{e4}a_5+\lambda_{e5}a_3=0$. So the first equation from (\ref{so.27}) implies
\begin{eqnarray}\label{so.28.2}
2e_4a_5+e_5a_3=0, \qquad 2\lambda_{e4}a_5+\lambda_{e5}a_3=0.
\end{eqnarray}
Similar analysis of the second equation from (\ref{so.27}) gives the equations
\begin{eqnarray}\label{so.30}
2e_3a_5+e_5a_4=0, \quad 2\lambda_{e3}a_5+\lambda_{e5}a_4=0.
\end{eqnarray}
The third equation from  (\ref{so.27}) does not imply new equations.

The resulting system (\ref{so.26})-(\ref{so.30}) is the typical case of degenerate Hamiltonian mechanics.

The auxiliary variables $\lambda_{e3}$, $e_3$, $\lambda_{e4}$,
$e_4$ and $\pi_{ea}$ are fixed by the algebraic equations in terms
of $\lambda_{e5}$. For the remaining variables we have the
differential equations
\begin{eqnarray}\label{so.31}
\dot e_5=\lambda_{e5},
\end{eqnarray}
\begin{eqnarray}\label{so.32}
\dot\omega^\mu=\frac{\partial H}{\partial\pi^\mu}-\frac{a_4}{2a_5}e_5\pi^\mu+\frac12 e_5\omega^\mu, \cr
\dot\pi^\mu=-\frac{\partial H}{\partial\omega^\mu}+\frac{a_3}{2a_5}e_5\omega^\mu+\frac12 e_5\pi^\mu,
\end{eqnarray}
as well as the  constraints (\ref{so.27}). We note that the
variable $\lambda_{e5}(\tau)$ cannot be determined with the
constraints, nor with the dynamical equations. As a consequence
(see  Eq. (\ref{so.31})), the variable $e_5$ turns out to be an
arbitrary function as well. Since $e_5(\tau)$ enters into the
equation for $\omega$ and $\pi$, their general solution contains,
besides the arbitrary integration constants, the arbitrary
function $e_5(\tau)$, $\omega=\omega(\tau, c^\mu, e_5(\tau))$.
Hence, all the basic variables has ambiguous dynamics. According
to the general theory \cite{dir, GT1, AAD1}, variables with
ambiguous dynamics do not represent the observable quantities.

So let us look for the variables with unambiguous dynamics. Consider the projection $J^{ij}(\tau)=f(\omega(\tau),
\pi(\tau))$. According the Proposition 4 and Eq. (\ref{so.8}), $J(\tau)$ lies on $\mathbb S$. Besides, $J(\tau)$
represents a solution to the problem (\ref{so.0})
\begin{eqnarray}\label{so.34}
\dot J^{\mu\nu}=\frac{\partial J^{\mu\nu}}{\partial\omega^\alpha}\dot\omega^\alpha+\frac{\partial
J^{\mu\nu}}{\partial\pi_\alpha}\dot\pi_\alpha= \qquad \qquad \cr \frac{\partial
J^{\mu\nu}}{\partial\omega^\alpha}\left(\frac{\partial H}{\partial\pi_\alpha}+e_3\pi^\alpha+\frac12
e_5\omega^\alpha\right)+ \cr \frac{\partial J^{\mu\nu}}{\partial\pi_\alpha}\left(-\frac{\partial
H}{\partial\omega^\alpha}-e_4\omega^\alpha-\frac12 e_5\pi^\alpha\right)= \cr \frac{\partial
J^{\mu\nu}}{\partial\omega^\alpha}\frac{\partial H}{\partial\pi_\alpha}-\frac{\partial
J^{\mu\nu}}{\partial\pi_\alpha}\frac{\partial H}{\partial\omega^\alpha}= \cr \left(\frac{\partial
J^{\mu\nu}}{\partial\omega^\alpha}\frac{\partial J^{\beta\gamma}}{\partial\pi^\alpha}- \frac{\partial
J^{\mu\nu}}{\partial\pi_\alpha}\frac{\partial J^{\beta\gamma}}{\partial\omega^\alpha}\right)\frac{\partial H}{\partial
J^{\beta\gamma}}= \cr \{J^{\mu\nu}, J^{\beta\gamma}\}_{PB}\frac{\partial H}{\partial J^{\beta\gamma}}=\{J^{\mu\nu},
J^{\beta\gamma}\}_{LPB}\frac{\partial H}{\partial J^{\beta\gamma}}= \cr \{J^{\mu\nu}, H(J)\}_{LPB}. \qquad \qquad
\end{eqnarray}
Hence we have obtained the desired result: any trajectory of the Hamiltonian flow of $H(J)$ on $\mathbb S$,
$J(\tau)|_{\mathbb S}$, is a projection of some trajectory $(\omega(\tau), \pi(\tau))$ of the variational problem
(\ref{so.25}), $\left. J(\tau)\right|_{\mathbb{S}}=f(\omega(\tau), \pi(\tau))$. In other words, trajectories of the
Lie-Poisson system (\ref{so.0}) lying on $\mathbb{S}\in\mathbb{R}^{\frac{n(n-1)}{2}}$, are obtained starting from the
variational problem (\ref{so.25}) formulated on $\mathbb{R}^{2n}$. \par

The invariance condition (\ref{so.6}) has been justified above from geometric point of view. We can also motivate it in
the framework of Dirac procedure. In the Hamiltonian formulation, equations (\ref{so.27}) appeared as the Dirac
constraints. So, we classify them in accordance to their algebraic properties with respect to the Poisson bracket. The
system of functions $T_j$ can be separated on two groups\footnote{Generally we need to consider an appropriate linear
combination of initial constraints \cite{GT1, AAD1}.}, $T=(G, K)$ in such a way that
\begin{eqnarray}
\{G, T\}\sim T, \qquad \{K, K\}=\triangle, \quad \det\triangle\ne 0.
\end{eqnarray}
That is Poisson bracket of $G$ with any constraint vanishes on the
constraint surface, while the Poisson brackets of $K$ form an
invertible matrix on the surface. In the Dirac terminology, the
set $G$ ($K$) is composed of first-class (second-class)
constraints. For the present case, any one of $T_a$ can be
separated as the first-class constraint. For example, the
combination $\tilde
T_5=T_5-\frac{a_4}{2a_5}T_3-\frac{a_3}{2a_5}T_4$ has vanishing
Poisson brackets with $T_3$ and $T_4$. Hence $T_3$ and $T_4$ form
the second-class pair while $\tilde T_5$ is the first-class
constraint.

Consistency of canonical quantization of a system with
second-class constraints implies replacement the Poisson by the
Dirac bracket, the latter is constructed with help of the
constraints. For the angular momenta it reads
\begin{eqnarray}
\{J^{\mu, \nu}, J^{\alpha\beta}\}_{DB}=\{J^{\mu, \nu},
J^{\alpha\beta}\}_{PB}- \cr \{J^{\mu, \nu}, K_a\}_{PB}\{K_a,
K_b\}_{PB}^{-1}\{K_b, J^{\alpha\beta}\}_{PB}.
\end{eqnarray}
If the constraints $K$ satisfy (\ref{so.6}), the second term on
the r. h. s. vanishes, and the Dirac bracket of $J^{\mu, \nu}$
reduces to the canonical Poisson bracket. So, as before, we are
dealing with the angular momentum algebra (\ref{so.1}).

First-class constraint $G$ imply a theory with local symmetry.
Generators of the symmetry are proportional to the first-class
constraints \cite{dir, GT1, AAD1}. Suppose that the first-class
constraints do not satisfy (\ref{so.6}), $\{G, J^{\mu, \nu}\}\ne
0$. This should imply that the variables $J^{\mu, \nu}$ are
affected by the local symmetry, $\delta J^{\mu, \nu}\sim\{G,
J^{\mu, \nu}\}\ne 0$. So, $J^{\mu, \nu}$ would be gauge
non-invariant variables, which is not of our interest now.

Above, we have specified the physical sector from analysis of equations of motion. The more traditional way to do this
consists of analysis of local symmetries of the formulation. For our case, presence of the first-class constraint
$\tilde T_5$ implies one-parametric local symmetry of the action (\ref{so.25}). This is just the local version of the
structure group transformations of section 3. For example, consider the infinitesimal transformation of Eq.
(\ref{sp.2}) with the local parameter $\beta(\tau)$. We absorb the factor $\tanh\sigma$ into $\beta$, then
\begin{eqnarray}\label{ls.0}
\delta\omega^\mu=-\beta\omega^\mu+\beta\pi^\mu\frac{\omega^2}{(\omega\pi)}, \cr
\delta\pi^\mu=-\beta\omega^\mu\frac{\pi^2}{(\omega\pi)}+\beta\pi^\mu.
\end{eqnarray}
By construction, the expression in square brackets of Eq. (\ref{so.25}) is invariant under the variation. Modulo to
total derivative, variation of the first term in (\ref{so.25}) can be presented as follows
\begin{eqnarray}\label{ls.1}
\delta(\pi\dot\omega)=-\frac{\beta\omega^2}{2(\omega\pi)}(\pi^2)^{.}-
\frac{\beta\pi^2}{2(\omega\pi)}(\omega^2)^{.}+\beta(\omega\pi)^{.}.
\end{eqnarray}
This can be cancelled by the following variations of auxiliary variables
\begin{eqnarray}\label{ls.2}
\delta e_3=\left(\frac{\beta\omega^2}{(\omega\pi)}\right)^{.}, \quad \delta
e_4=\left(\frac{\beta\pi^2}{(\omega\pi)}\right)^{.}, \cr \delta e_5=-2(\beta)^{.}, \quad \delta\lambda_{ea}=(\delta
e_a)^{.}
\end{eqnarray}
Hence the equations (\ref{ls.0}) and (\ref{ls.2}) represent the spin-plane local symmetry of the action (\ref{so.25}).
We have verified that the finite transformation (\ref{sp.1}), being accompanied by a complicated transformation law of
$e_a$, represents a local symmetry as well.

\section{Lagrangian action}
For the frozen spin, the initial Hamiltonian (\ref{so.22}) is a
scalar function of $J$, that is some combination of Casimir
operators. As we have mentioned above, this implies $H=H(\pi^2,
\omega^2, (\omega\pi))$. This allows us to use the constraints
$T_a=0$ in those terms of Eq. (\ref{so.28}) which contain
derivative of the Hamiltonian. Let us denote
\begin{eqnarray}\label{la.1}
H_{\pi\pi}=\left.\frac{\partial H}{\partial\pi^2}\right|_{T_a=0}, \quad H_{\omega\pi}=\left.\frac{\partial
H}{\partial(\omega\pi)}\right|_{T_a=0}, \cr H_{\omega\omega}=\left.\frac{\partial H}{\partial\omega^2}\right|_{T_a=0}.
\qquad \qquad
\end{eqnarray}
Then Eq. (\ref{so.28}) is equivalent to
\begin{eqnarray}\label{la.2}
\dot
\omega^\mu=2H_{\pi\pi}\pi^\mu+H_{\omega\pi}\omega^\mu+e_3\pi^\mu+\frac12
e_5\omega^\mu,
\end{eqnarray}
\begin{eqnarray}\label{la.3}
\dot\pi^\mu=-2H_{\omega\omega}\omega^\mu-H_{\omega\pi}\pi^\mu-e_4\omega^\mu-\frac12
e_5\pi^\mu.
\end{eqnarray}
They follow from the Hamiltonian action
\begin{eqnarray}\label{la.4}
S_H=\int d\tau ~ \pi\dot\omega-\left[H_{\pi\pi}\pi^2+
H_{\omega\pi}(\omega\pi)+H_{\omega\omega}\omega^2
+\frac{e_a}{2}T_a\right].
\end{eqnarray}
We solve Eq. (\ref{la.2}) with respect to $\pi^\mu$ and substitute
the result into Eq. (\ref{la.4}). This gives the Lagrangian action
\begin{eqnarray}\label{la.5}
S=\int d\tau \frac{1}{2\tilde e_3}(D\omega^\mu)^2-\frac12\tilde e_4\omega^2-\frac{\tilde e_a}{2}a_a, \quad a=3, 4, 5.
\end{eqnarray}
We have denoted
\begin{eqnarray}\label{la.6}
D\omega^\mu=\dot\omega^\mu-\frac12\tilde e_5\omega^\mu.
\end{eqnarray}
\begin{eqnarray}\label{la.7}
\tilde e_3=e_3+2H_{\pi\pi}, \quad \tilde e_4=e_4+2H_{\omega\omega}, \quad \tilde e_5=e_5+2H_{\omega\pi},
\end{eqnarray}
We point out that the coefficients $H_{\pi\pi}$, $H_{\omega\pi}$ and $H_{\omega\omega}$ can be absorbed by $e_a$, that
is the spin surface of a frozen spin does not admit non trivial selfinteraction.

Owing to the expressions (\ref{la.2}) and (\ref{la.6}) we can write
\begin{eqnarray}\label{la.8}
\pi^\mu=\frac{1}{\tilde e_3}D\omega^\mu.
\end{eqnarray}
We substitute this expression into Eqs. (\ref{ls.0}) and (\ref{ls.2}), this gives local symmetry of the Lagrangian
action (\ref{la.5})
\begin{eqnarray}\label{la.9}
\delta\omega^\mu=-\beta K^\mu{}_\nu\omega^\nu, \quad \delta\tilde e_5=-2\dot\beta, \qquad \quad \cr \delta\tilde
e_3=\left(\frac{\beta\tilde e_3\omega^2}{(\omega D\omega)}\right)^., \quad \delta\tilde e_4=\left(\frac{\beta (D\omega
D\omega)}{\tilde e_3(\omega D\omega)}\right)^..
\end{eqnarray}
where
\begin{eqnarray}\label{la.10}
K^\mu{}_\nu=\delta^\mu{}_\nu-\frac{D\omega^\mu\omega_\nu}{(\omega D\omega)}, \cr \omega_\mu K^\mu{}_\nu=0, \quad
K^\mu{}_\nu D\omega^\nu=0.
\end{eqnarray}

\section{Conclusion}
In this work we have formulated variational problem for Hamiltonian system (\ref{so.0}) with $so(k, m)$ Lie-Poisson
bracket (\ref{so.1}) which propagate on $2n-4$\,-dimensional spin surface defined by Eq. (\ref{so.8}). Our main
motivation for restriction the dynamics on the surface is that for the cases $so(3)$, $so(1, 3)$ and $so(2, 3)$, this
describes dynamics of semiclassical spin, see \cite{AAD2, AAD3, AAD7, AAD4}.

To formulate the variational problem according the standard prescription (\ref{so.0.1}), we embed the spin surface into
phase-space with canonical Poisson bracket. The embedding procedure can be resumed as follows.

First, we have identified the spin surface with base of $2n-3$\,-dimensional spin fiber bundle defined by Eqs.
(\ref{so.7}) and (\ref{so.9}). Structure group has been described in section 3.

Second, the fiber bundle has been embedded as a surface into $2n$\,-dimensional phase space equipped with the canonical
Poisson bracket (\ref{so.2}). The projection map (\ref{so.3}) implies that the Lie-Poisson bracket (\ref{so.1}) is
generated by the Poisson one, see Eq. (\ref{so.5.1}).

Further, we treat the embedding as imposition of constraints on the phase space, and look for the action functional
which implies the constraints. This results in the Hamiltonian action functional (\ref{so.25}). We have verified that
this implies the constraints (\ref{so.7}) as well as the desired Hamiltonian equations (\ref{so.0}). The corresponding
Lagrangian action is given by Eq. (\ref{la.5}).

We point out that the constraints fix values of $SO(n)$ Casimir operators, which implies the possibility of unambiguous
canonical quantization. Appearance the first-class constraint $T_5=(\omega\pi)+a_5=0$ reflects invariance of the action
under local (gauge) symmetry. The symmetry is just the structure group transformation acting independently at each
instance of time. The spin-plane local symmetry play the fundamental role, determining the gauge-invariant variables
and, at the end, physical sector of the spinning particles proposed in \cite{AAD2, AAD3, AAD7, AAD4}.

\appendix
\section{Phase space associated with a linear representation of Lie algebra}
Let $\{e^a, e^b\}=c^{ab}{}_ce^c$ be Lie algebra with generators
$e^a$, and $\varphi: e^a\rightarrow\varphi(e^a)$ be a linear
representation
\begin{eqnarray}\label{ap.2}
\varphi\left(\{e^a,
e^b\}\right)=\varphi(e^a)\varphi(e^b)-\varphi(e^b)\varphi(e^a),
\end{eqnarray}
of the algebra on a vector space with the coordinates
$\omega^\alpha$,
\begin{eqnarray}\label{ap.3}
\omega^\alpha\rightarrow\omega'^\alpha=(\varphi^a)^\alpha{}_\beta\omega^\beta,
\end{eqnarray}
where $(\varphi^a)^\alpha{}_\beta$ stands for the matrix which
represents the transformation $\varphi(e^a)$. Eq. (\ref{ap.2})
implies that the matrices obey the same algebra as $e^a$
\begin{eqnarray}\label{ap.4}
\varphi^a\varphi^b-\varphi^b\varphi^a=c^{ab}{}_c\varphi^c.
\end{eqnarray}
To arrive at the Lie-Poisson bracket for variables $z^a$
\begin{eqnarray}\label{ap.1}
\{z^a, z^b\}=c^{ab}{}_cz^c,
\end{eqnarray}
starting from the canonical Poisson structure, we introduce phase
space with the coordinates $(\omega^\alpha, \pi_\beta)$ equipped
with Poisson bracket
\begin{eqnarray}\label{ap.5}
\{\omega^\alpha, \pi_\beta\}_{PB}=\delta^\alpha{}_\beta,
\end{eqnarray}
and use the representation (\ref{ap.2}) to construct the quantities
\begin{eqnarray}\label{ap.6}
z^a=(\varphi^a)^\alpha{}_\beta\omega^\beta\pi_\alpha.
\end{eqnarray}
As a consequence of (\ref{ap.4}), their Poisson bracket generates
(\ref{ap.1})
\begin{eqnarray}\label{ap.7}
\{z^a, z^b\}_{PB}=(\varphi^a\varphi^b-\varphi^b\varphi^a)^\alpha{}_ \beta\omega^\beta\pi_\alpha= \cr
c^{ab}{}_c(\varphi^c)^\alpha{}_\beta\omega^\beta\pi_\alpha=c^{ab}{}_cz^c.
\end{eqnarray}

In particular, any Lie algebra admits adjoint representation
defined by the map $e^a\rightarrow -(c^a)^b{}_c$. Then Poisson
brackets of the quantities
\begin{eqnarray}\label{ap.8}
z^a=-c^{ab}{}_c\omega^c\pi_b,
\end{eqnarray}
generate the Lie-Poisson bracket (\ref{ap.1})
\begin{eqnarray}\label{ap.9}
\{z^a, z^d\}_{PB}=c^{ab}{}_cc^{de}{}n\{\omega^c\pi_b,
\omega^n\pi_e\}_{PB}= \cr
-(c^{ab}{}_cc^{de}{}_b+c^{db}{}_cc^{ea}{}_b)\omega^c\pi_e=
c^{eb}{}_cc^{ad}{}_b\omega^c\pi_e=c^{ad}{}_bz^b,
\end{eqnarray}
In this computation we have used Jacobi identity for structure
constants.

In section 2 we use the phase space associated with vector representation $so(n)$ in $n$\,-dimensional real space with
the coordinates $\omega^i$, $\omega'^i=\epsilon^{ij}\omega^j$. Here $\epsilon^{ij}$ is antisymmetric matrix. The
corresponding matrix realization for generators is
\begin{eqnarray}\label{ap.10}
(\varphi^{ij})^k{}_n=\frac{1}{2}(\delta^{ik}\delta^j{}_n-\delta^{jk}\delta^i{}_n), ~
\omega'{}^k=\epsilon_{ij}(\varphi^{ij})^k{}_n\omega^n.
\end{eqnarray}
According to the prescription given above, we introduce
$2n$\,-dimensional phase space with coordinates $\omega^i$,
$\pi_j$, equipped with the Poisson bracket
\begin{eqnarray}\label{ap.11}
\{\omega^i, \pi_j\}_{PB}=\delta^i{}_j,
\end{eqnarray}
and define the inner angular momentum according to Eq.
(\ref{ap.6})
\begin{eqnarray}\label{ap.12}
J^{ij}=4(\varphi^{ij})^k{}_n\omega^n\pi_k\equiv
2(\omega^i\pi^j-\omega^j\pi^i).
\end{eqnarray}
Poisson bracket of these quantities coincides with
$so(n)$\,-Lie-Poisson bracket (\ref{so.1}).

\section{Identification of the standard fiber}
Here we describe standard fiber $\mathbb{F}_J$ of the spin fiber
bundle (\ref{so.9}). $\mathbb{F}_J$ is preimage of a point
$J\in\mathbb{S}$, $\mathbb{F}_J=f^{-1}(J)$. We identify the vector
$(\omega, \pi)\in\mathbb{T}^{2n-3}$ with the pair of orthogonal
vectors of $\mathbb{R}^n$
\begin{eqnarray}\label{so.9.1.1}
\vec\pi^2=-a_3, \quad \vec\omega^2=-a_4, \quad (\vec\omega, \vec\pi)=-a5.
\end{eqnarray}
Square of Casimir operator is $J^2=8[a_3a_4-a_5^2]$. Since we are interested in the case $J^2\ne 0$, the only
restriction on the numbers $a_i$ is $a_3a_4-a_5^2\ne 0$. Note also that this implies
\begin{eqnarray}\label{ap.12.3}
\vec \omega\ne\lambda\vec\pi.
\end{eqnarray}

We state that $\mathbb{F}_J$ is composed by all the pairs obtained from $(\vec\omega, \vec\pi)$ by rotations in the
plane of these vectors.

To confirm this, we need to find all solutions of the system $2\omega^{[i}\pi^{j]}=J^{ij}$ with given right hand side.
Let $(\omega_0, \pi_0)\in\mathbb{F}_J$. Having in mind the identification (\ref{so.9.1.1}), let us take coordinates in
$\mathbb{R}^n$ such that the first two basic vectors of the system lie on the plane of vectors $\vec\omega_0$ and
$\vec\pi_0$. In this system they are
\begin{eqnarray}\label{so.9.2}
\begin{array}{ccccc}
(\omega^2,& \omega^2,& 0, &\ldots ,& 0)\\
(\pi^1,& \pi^2,& 0,& \ldots ,& 0)
\end{array}
\end{eqnarray}
Hence our task is to solve the system
\begin{eqnarray}\label{so.9.3}
2\omega^{[i}\pi^{j]}=J^{ij}, \quad \mbox{where} \quad J^{12}=-J^{21}=2\sqrt{a_3a_4-a_5^2}, \cr J^{ij}=0. \qquad \qquad
\quad
\end{eqnarray}
Evidently, all the pairs obtained from $(\vec\omega, \vec\pi)$ by
rotations in the plane of these vectors belong to $\mathbb{F}_J$.
Let us show that they are the only elements of $\mathbb{F}_J$.
Observe that the system (\ref{so.9.3}) is the statement on values
of minors
\begin{eqnarray}\label{so.9.5}
m(ij)=\det\left(
\begin{array}{ccccc}
\omega^i& \omega^j\\
\pi^i& \pi^j
\end{array}
\right).
\end{eqnarray}
of the matrix
\begin{eqnarray}\label{so.9.6}
\left(
\begin{array}{ccccc}
\omega^1,& \omega^2,& \omega^3, &\ldots ,& \omega^n\\
\pi^1,& \pi^2,& \pi^3,& \ldots ,& \pi^n
\end{array}
\right).
\end{eqnarray}
First, we demonstrate that (\ref{so.9.3}) implies $\omega^3=\pi^3=0$.

Suppose that (\ref{so.9.3}) has a solution with $\omega^3\ne 0$. Consider the equations from (\ref{so.9.3}) which
correspond to the minors
\begin{eqnarray}\label{so.9.7}
m(23)=\omega^2\pi^3-\omega^3\pi^2=0,
\end{eqnarray}
\begin{eqnarray}\label{so.9.8}
m(13)=\omega^1\pi^3-\omega^3\pi^1=0,
\end{eqnarray}
\begin{eqnarray}\label{so.9.9}
m(1j)=\omega^1\pi^j-\omega^j\pi^1=0, \quad j=4, 5, \ldots , n.
\end{eqnarray}
They imply
\begin{eqnarray}\label{so.9.10}
\pi^2=\frac{\pi^3}{\omega^3}\omega^2,
\end{eqnarray}
\begin{eqnarray}\label{so.9.11}
\pi^1=\frac{\pi^3}{\omega^3}\omega^1.
\end{eqnarray}
Use (\ref{so.9.11}) in (\ref{so.9.9})
\begin{eqnarray}\label{so.9.12}
\omega^1(\pi^j-\frac{\pi^3}{\omega^3}\omega^j)=0.
\end{eqnarray}
If $\omega^1=0$, than $m(13)=\omega^1\pi^3-\omega^3\pi^1=0$ implies $\pi^1=0$, then $m(12)=0$. This is in contradiction
with $m(12)=\sqrt{a_3a_4}\ne 0$. If $\omega^1\ne 0$, Eq. (\ref{so.9.12}) implies
$\pi^j=\frac{\pi^3}{\omega^3}\omega^j$, $j=4, 5, \ldots , n$. Then $\omega^3\ne 0$ implies
$\vec\pi=\frac{\pi^3}{\omega^3}\vec\omega$, in contradiction with Eq. (\ref{ap.12.3}). Thus (\ref{so.9.3}) implies
$\omega^3=0$.

Having in mind $\omega^3=0$, consider the following equations from (\ref{so.9.3})
\begin{eqnarray}\label{so.9.13}
m(13)=\omega^1\pi^3=0,
\end{eqnarray}
\begin{eqnarray}\label{so.9.14}
m(23)=\omega^2\pi^3=0,
\end{eqnarray}
\begin{eqnarray}\label{so.9.15}
m(3j)=-\omega^j\pi^3=0, \quad j=4, 5, \ldots , n.
\end{eqnarray}
If $\pi^3\ne 0$, then $\vec\omega=0$. So $J^{ij}=0$, which is in contradiction with $J^2\ne 0$. Hence $\pi^3=0$.

We continue the process, obtaining $\omega^i=\pi^i=0$, $i=3, 4, \ldots , n$. So the only solutions of the system
(\ref{so.9.3}) are the vectors
\begin{eqnarray}\label{so.9.16}
\begin{array}{ccccc}
(\omega^1,& \omega^2,& 0, &\ldots ,& 0)\\
(\pi^1,& \pi^2,& 0,& \ldots ,& 0)
\end{array} \mbox{where} \left\{
\begin{array}{c}
\omega^{[1}\pi^{2]}=\sqrt{a_3a_4}, \\
(\omega^1)^2+(\omega^2)^2=-a_4, \\
(\pi^1)^2+(\pi^2)^2=-a_3, \\
\omega^1\pi^1+\omega^2\pi^2=-a_5
\end{array}
\right.
\end{eqnarray}
as it has been stated.

\acknowledgments This work has been supported by the Brazilian foundation FAPEMIG.

\end{document}